\documentclass[aps,prl,superscriptaddress,floatfix]{revtex4}
\usepackage{amsfonts}
\usepackage{epsfig}
\usepackage{amsmath,amssymb}
\usepackage{times}
\usepackage{xcolor}
\usepackage{multirow}

\usepackage{graphicx}
\usepackage{color}
\usepackage[labelfont=bf,labelsep=period,justification=raggedright]{caption}
\date{}

\begin{document}
\title{Core-like groups result in invalidation of identifying super-spreader by k-shell decomposition}

\author{Ying Liu}
\affiliation{Web Sciences Center, University of Electronic
Science and Technology of China, Chengdu 610054, China}
\affiliation{School of Computer Science, Southwest Petroleum University,
Chengdu 610500, China}

\author{Ming Tang\footnote{Correspondence to: tangminghuang521@hotmail.com}}
\affiliation{Web Sciences Center, University of Electronic
Science and Technology of China, Chengdu 610054, China}

\author{Tao Zhou}
\affiliation{Web Sciences Center, University of Electronic
Science and Technology of China, Chengdu 610054, China}
\affiliation{Big Data Research Center, University of Electronic
Science and Technology of China, Chengdu 610054, China}

\author{Younghae Do}
\affiliation{Department of Mathematics, Kyungpook National
University, Daegu 702-701, South Korea}

\begin{abstract}
Identifying the most influential spreaders is an important issue in understanding and controlling spreading processes on complex networks. Recent studies showed that nodes located in the core of a network as identified by the $k$-shell decomposition are the most influential spreaders. However, through a great deal of numerical simulations, we observe that not in all real networks do nodes in high shells are very influential: in some networks the core nodes are the most influential which we call true core, while in others nodes in high shells, even the innermost core, are not good spreaders which we call core-like group. By analyzing the $k$-core structure of the networks, we find that the true core of a network links diversely to the shells of the network, while the core-like group links very locally within the group. For nodes in the core-like group, the $k$-shell index cannot reflect their location importance in the network. We further introduce a measure based on the link diversity of shells to effectively distinguish the true core and core-like group, and identify core-like groups throughout the networks. Our findings help to better understand the structural features of real networks and influential nodes.
\end{abstract}

%\pacs{05.45.-a,89.75.-k}
\date{\today}
\maketitle

%While the networks supporting the two types of spreading process,
%one communication layer and another the layer of physical contact, share
%the same set of nodes, in a realistic situation the detailed connecting
%structures of the layers can be quite different, leading to a complex
%double-layer network with asymmetrical interactions. \\

The most influential nodes can maximize the speed and scope of information spreading compared with other nodes in a network{~\cite{kempe2003}}. Locating these influential nodes is important in improving the use of available resources~\cite{Gallos2007} and controlling the spread of information{~\cite{lloyd2005}}. A critical issue is how to determine and distinguish the spreading capability of a node. Centrality is usually used to measure the relative importance of nodes within the network, such as degree centrality~\cite{freeman1978}, betweenness centrality~\cite{freeman1977}, closeness centrality~\cite{sabi1966}, eigenvector centrality~\cite{bonacich2001}, PageRank centrality~\cite{page1998} and its variance~\cite{kleinberg1999}. Nodes with high centrality are considered more influential in the spreading process~\cite{geoffrey2006,linyuan2006,kitsak2010,chen2012,macdonald2012}. Among these measures, degree centrality is a simple and effective way, although it is based only on local link  information{~\cite{borge20122,Tanaka2012}}. The merit of degree is challenged by a recent study~\cite{kitsak2010}, in which the authors pointed out that the most influential spreaders do not correspond to the nodes with largest degree, but are those located in the core of the network as identified by the $k$-shell decomposition~\cite{bolobas1984}. This means the higher coreness of a node, the more influential it is in the spreading dynamics.

The $k$-shell decomposition decomposes a network into hierarchically ordered shells by recursively pruning the nodes with degree less than current shell index (see \textbf{Methods} for details ). This procedure assigns each node with an index $k_S$, representing its coreness in the network. A large $k_S$ value means a core position in the network, while a small $k_S$ value defines the periphery of the network{~\cite{kitsak2010}}. Because of a low computational complexity of $O(N+E)${~\cite{batagelj2003}}, where $N$ is the network size and $E$ is the number of edges in the network, this method is extensively used for large-scale network analysis. Generally speaking, it is used to efficiently visualize the structure of large-scale networks~\cite{hamelin2005,shai2007}, analyze the core structure of networks~\cite{dorogovtsev2006,hamelin2008,colomer2013,holme2005,zhang2008}, and capture the essential structural properties of real networks~\cite{hebert2013}. Since the publication of Ref.~\cite{kitsak2010}, the coreness is widely used to quantify the importance of a node in a spreading process~\cite{Castellano2012,Senpei2013,Fu2014}. For example, in a economic crisis network, nodes with the highest coreness are most likely to spread a crisis globally~\cite{garas2010}, while in a rumor spreading, nodes with high coreness act as firewalls to prevent the diffusion of a rumor to the whole system~\cite{borge2012}. Even nodes with low coreness are considered as bridge elements, which can effectively control the disease in small world networks through an acquaintance-based vaccination strategy~\cite{reppas2012}. Many works extended the $k$-shell decomposition method, either modify it for a better ranking~\cite{zeng2013,brown,murtra2013,liu2013,bae2014} or generalize it to weighted networks~\cite{garas2012,marius2013}, dynamical networks~\cite{miorandi2010} and multiplex networks~\cite{Azimi2014}.

In all these studies, the $k$-shell decomposition is used as a powerful tool to analyze the network structure and identify important nodes. Despite its effectiveness, researchers have noticed that the $k$-shell method has some defects~\cite{Senpei2013,Senpei2014}. For example, when there is a lack of the complete network structure, one can not apply the $k$-shell decomposition to the network. In tree structure and BA model network, the capability of finding influential spreaders is limited due to the low resolution of $k_S$ index. Here, from the perspective of spreading efficiency of cores that have been identified by the $k$-shell decomposition, we study on whether in different real networks do the core nodes have a higher spreading influence than other nodes. In a common belief, it is. But through intensive computer simulations, we find that it is not the case. In some networks core nodes have the largest spreading efficiency, while in others core nodes have relatively low spreading efficiency.  What is the reason for the obvious distinct results? No work has focused on this question to our knowledge. In Refs.~\cite{kitsak2010,bae2014}, the authors pointed out that the performance of centrality measure relates somehow to the infection probability when evaluating the spreading capability of nodes. We find that although the infection probability will cause some fluctuations, the specific structure of real networks is the origin of the distinct performance of coreness in predicting spreading efficiency: in the first case, the core of a network has a link diversity to other shells of the network, while in the latter case the core is linked very locally. We respectively call them true core and core-like group. Then, we propose a measure of information entropy to locate core-like groups in real networks. These findings will help us in understanding the real network structures.
% Results and Discussion can be combined.

\section*{Results}
We first calculate the imprecision of coreness and degree in identifying influential spreaders and discover the true core and core-like group in real networks. We then analyze the structural features of the true core and core-like group and uncover their difference. Finally we successfully locate the core-like groups throughout the network by defining a measure of link entropy.

\textbf{Calculating the imprecision function of coreness and degree in the SIR spreading process.} We use a classic Susceptible-Infected-Recovered (SIR) spreading model to simulate the spreading process~\cite{anderson1992,moreno2002}, and record the spreading capability (or spreading efficiency) of each node, which is defined as the average size of infected population $M$ for each node as spreading origin (see \textbf{Methods} for details). To evaluate whether the structural centrality of coreness is an effective index to measure the spreading capability of nodes compared with degree, we calculate the imprecision function $\epsilon_{k_S}(p)$ and $\epsilon_{k}(p)$ proposed in Ref.~\cite{kitsak2010}. The imprecision function is defined as
\begin{equation}\label{imprecision}
\varepsilon_{k_S(k)}(p)=1-\frac{M_{k_S(k)}(p)}{M_{eff}(p)},
\end{equation}
where $p$ is the fraction of network size $N$ ($p\in[0,1]$), $M_{k_S(k)}(p)$ and $M_{eff}(p)$ are the average spreading efficiencies of $pN$ nodes with highest coreness $k_S$ (degree $k$) values and largest spreading efficiency, respectively. This function quantifies how close to the optimal spreading is the average spreading of the $pN$ nodes with largest $k_S$ ($k$) values. The smaller the $\varepsilon_{k_S(k)}$ value, the more accurate the $k_S(k)$ index is a measure to identify the most influential spreaders.

\begin{figure}[!ht]
\begin{center}
\epsfig{file=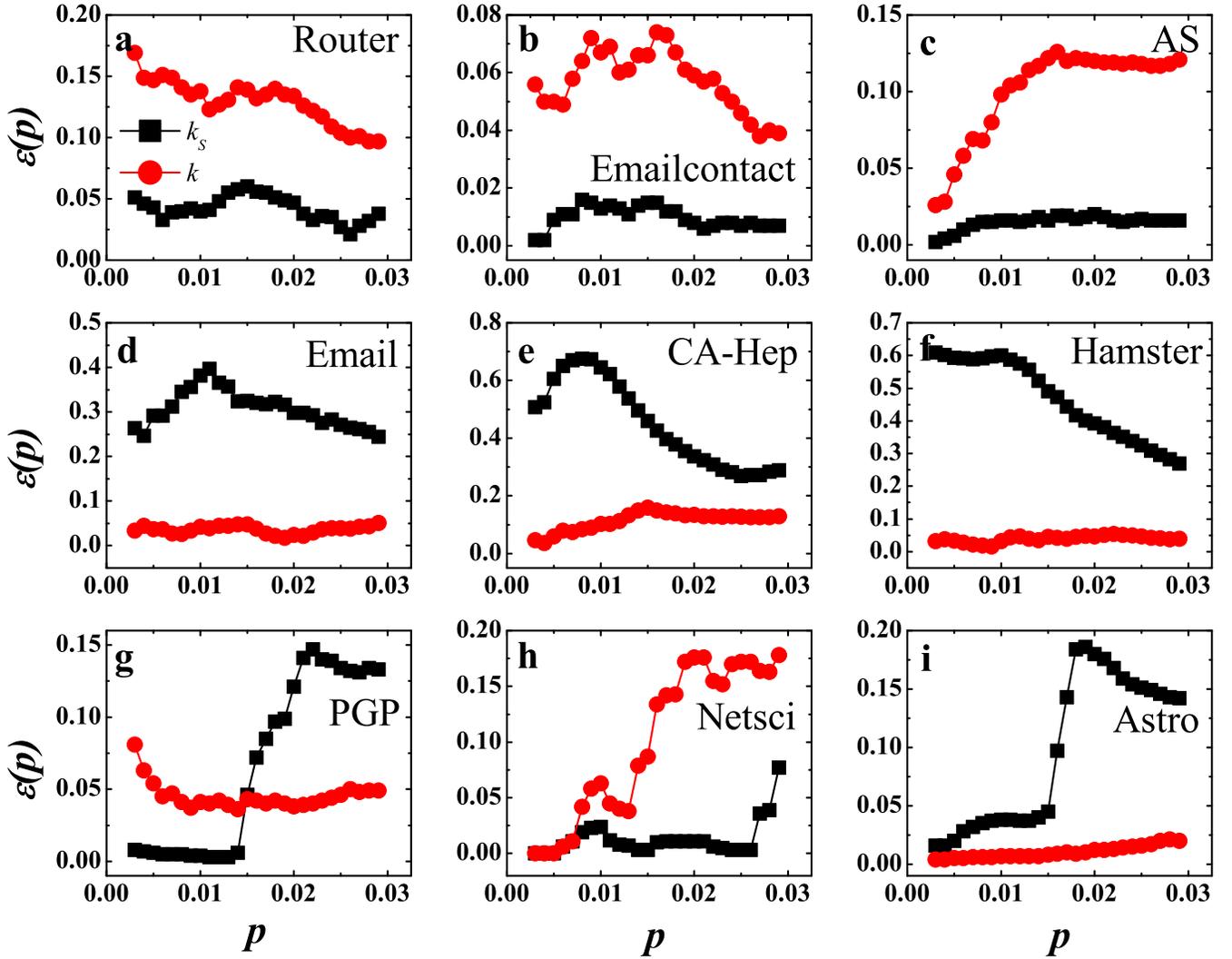,width=1\linewidth}

\caption{\textbf{The imprecision of $k_S$ and $k$ as a function of $p$ for nine real networks.} The $k_S$ imprecision (black squares) and $k$ imprecision (red circles) are compared in each network. $p$ is the proportion of nodes calculated, ranging from 0.003 to 0.029.  See Fig.S1 for large $p$ plots in SI.}
\label{figure1}
\end{center}
\end{figure}

The imprecision functions of nine real networks are shown in Fig.~\ref{figure1}. Contrary to common belief, the coreness $k_s$ does not perform consistently well in all networks. We divide them into three groups. In Router, Emailcontact and AS networks, the $k_S$ imprecision is lower than the $k$ based method. In Fig.~\ref{figure1} (a)-(c), the imprecision $\epsilon_{k_S}(p)$ is very low, under $0.06$ in the demonstrated range of $ 0.003\leqslant p \leqslant 0.029$, and is much lower than $\epsilon_{k}(p)$. This means the coreness predicts the outcome of spreading more reliably than degree. However, the imprecision $\epsilon_{k_S}(p)$ for the next three networks (i.e., Email, CA-Hep and Hamster) is much higher than the imprecision $\epsilon_{k}(p)$. In Fig.~\ref{figure1} (d)-(f), the values of $\epsilon_{k_S}(p)$ is above $0.2$ for all the three networks, and is much higher than $\epsilon_{k}(p)$. This is completely contrary to the case of the first three networks. As for the last three networks of PGP, Netsci and Astro networks, things are more complicated shown in Fig.~\ref{figure1} (g)-(i). In PGP, the $k_S$ method acts better than $k$ when $p<0.015$. Then there is a sudden rise in $\epsilon_{k_S}(p)$ and it becomes higher than the imprecision of degree. In Netsci, when $p\leq0.026$, the imprecision of $k_S$ is much lower than that of $k$. There is a fast rise of the $\epsilon_{k_S}(p)$ at $p=0.027$, and at $p=0.05$ the $k_S$ imprecision exceeds $k$ imprecision (see Fig. S1 in Supporting Information (SI) for large $p$ plots). In Astro, the sudden rise of $k_S$ imprecision occurs at $p=0.015$ and the value of $\epsilon_{k_S}(p)$ goes up to around $0.18$. This indicates a complex performance of coreness as a measure of spreading efficiency.

\begin{figure}[!ht]
\begin{center}
\epsfig{file=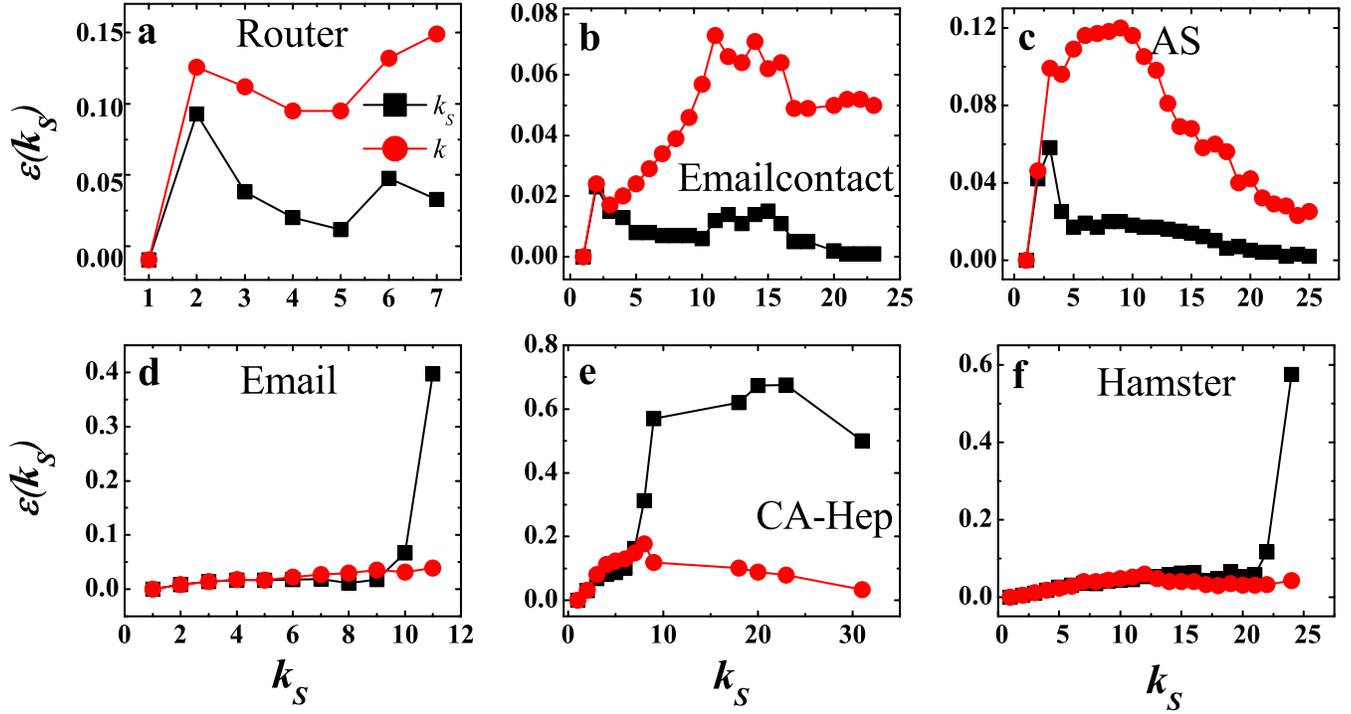,width=1\linewidth}

\caption{\textbf{The imprecision of $k_S$ and $k$ as a function of $k_S$ for six real networks.} The $k_S$ imprecision (black squares) and $k$ imprecision (red circles) are compared in each network. Each square represents the $k_S$ imprecision of nodes in $k_S$-core, and each circle represents the $k$ imprecision of $n$ highest degree nodes, where $n$ equals to the number of nodes in $k_S$-core. $k_S$ is an integer representing the shell index, ranging from the smallest $k_S$ value to the largest $k_S$ value in the network.}
\label{figure2}
\end{center}
\end{figure}

\textbf{Discovering true core and core-like group in real networks.} To find out the reason for the distinct performance of coreness in predicting spreading efficiency is the origin of our research interest in this paper. In the following, we first explore the structural characteristics of the first two groups of networks, and then explain the performance of coreness in the last three networks.
As we know, the $k$-shell decomposition tends to assign many nodes with identical $k_S$ value, although their spreading capabilities may be different. When we calculate the imprecision function at a certain $p$, nodes with the same $k_S$ value are chosen randomly. This will cause some fluctuation in the $k_S$ imprecision curve (fraction of nodes in high shells is shown in the SI table S1). Given this fluctuation, we change to calculate \begin{equation}
\varepsilon_{k_S}(k_S)=1-\frac{M_{k_S}(k_S)}{M_{eff}(k_S)},
 \end{equation}
where $M_{k_S}(k_S)$ is the average spreading efficiency of the nodes with coreness $k_S'\geq k_S$ (nodes in $k_S$-core), and $M_{eff}(k_S)$ is the average spreading efficiency of $n$ nodes with highest spreading efficiency, where $n$ equals to the number of nodes with  coreness $k_S'\geq k_S$. To compare with $k$ performance, we have \begin{equation}
\varepsilon_{k}(k_S)=1-\frac{M_{k}(k_S)}{M_{eff}(k_S)},
\end{equation}
where $M_{k}(k_S)$ is the average spreading efficiency of $n$ nodes with highest degree, and n is as above. The imprecision of $k_S$ is supposed to be low if the nodes in high shells are efficient spreaders. The results are shown in Fig.~\ref{figure2}. In the first three networks (a)-(c), the $\varepsilon_{k_S}(k_S)$ is very low and much lower than the imprecision of degree $\varepsilon_{k}(k_S)$ for large $k_S$, which means most of nodes in high shells (shells with large $k_S$ value) are efficient spreaders. In the next three networks (d)-(f), the $\varepsilon_{k_S}(k_S)$ is much higher than the $\varepsilon_{k}(k_S)$ for the innermost core (the shell with the maximum $k_S$ value), and the absolute value is even greater than $0.4$, which means many nodes in the innermost core are not influential spreaders. From the perspective of dynamic spreading, we call the innermost core of the first three networks a \emph{true core}, and presumably call that of the other three networks a \emph{false core}, or \emph{core-like group}. This poor $k_S$ performance is obviously different from the fluctuation of imprecision caused by the resolution of $k_S$ index we mentioned above.

\textbf{Exploring the cause of poor coreness imprecision from structural features.} In order to find out the reason for the poor performances of coreness in the spreading process, we first look into the structural properties of the studied real networks. The features of the studied real networks are listed in Table~\ref{tab:basiccharacteristic}. From Table~\ref{tab:basiccharacteristic}, we see that the degree heterogeneity $H_{k}$ of the first group is sufficiently larger than that of the second group. The degree heterogeneity is defined as $H_{k}=\langle k^{2} \rangle/\langle k \rangle^{2}$ that evaluates the heterogeneity of degree sequence of a network, where $\langle k^{2} \rangle$ and $\langle k \rangle$ are the second moment and first moment of degree respectively. In addition, the degree assortativity $r$ of the first group is negative, which implies that nodes of large degrees are inclined to connecting to nodes of small degrees. As nodes in high shells always have large degrees and nodes in low shells (shells with small $k_S$ value) have small degrees, negative assortativity implies a good connection between high shell nodes and low shell nodes. On the contrary, the assortativity of the second group is positive or close to zero, which implies nodes of large degrees are inclined to connect to each other or connect randomly.

\begin{table*}[!ht]
\caption{\textbf{Properties of the real networks studied in this work.} Structural properties include number of nodes ($N$), number of edges ($E$), average degree ($\langle k \rangle$), maximum degree ($k_{max}$), degree heterogeneity ($H_k$), degree assortativity ($r$), clustering coefficient ($C$), maximum $k_S$ index ($k_{Smax}$), epidemic threshold ($\lambda_{c}$), infection probability in the SIR spreading in the main text ($\lambda$). }\centering

\begin{tabular}{cccccccccccc}
\hline
\hline
\textbf{Network} & \textbf{$N$} & \textbf{$E$} & \textbf{$\langle k \rangle$} & \textbf{$k_{max}$} & \textbf{$H_{k}$} & \textbf{$r$} & \textbf{$C$} & \textbf{$k_{Smax}$}& \textbf{$\lambda_{c}$} & \textbf{$\lambda$}\\
\hline
%\rowcolor{mygray}
Router  &5022 &6258 &2.5 &106 &5.503 &-0.138 &0.012 &7 &0.08 &0.27\\
%\hline
%\rowcolor{mygray}
Emailcontact  &12625 &20362 &3.2 &576 &34.249 &-0.387 &0.109 &23 &0.01 &0.10\\
%\hline
%\rowcolor{mygray}
AS  &22963 &48436 &4.2 &2390 &61.978 &-0.198 &0.230  &25 &0.004 &0.13\\
\hline
%\rowcolor{mypink}
Email  &1133 &5451 &9.6 &71 &1.942 &0.078 &0.220 &11 &0.06 &0.08\\
%\hline
%\rowcolor{mypink}
CA-Hep &8638 &24806 &5.7 &65 &2.261 &0.239 &0.482 &31 &0.08 &0.12\\
%\hline
%\rowcolor{mypink}
Hamster  &2000 &16097 &16.1 &273 &2.719 &0.023 &0.540 &24 &0.02 &0.04\\
\hline
%\rowcolor{mycyan}
PGP  &10680 &24340 &4.6 &206 &4.153 &0.240 &0.266 &31 &0.06 &0.19\\
%\hline
%\rowcolor{mycyan}
Netsci  &379 &914 &4.8 &34 &1.663 &-0.082 &0.741 &8 &0.14 &0.30\\
%\hline
%\rowcolor{mycyan}
Astro  &14845 &119652 &16.1 &360 &2.820 &0.228 &0.670 &56 &0.02 &0.05\\
\hline
\hline
\end{tabular}
%\begin{flushleft}
%\end{flushleft}
\label{tab:basiccharacteristic}
\end{table*}

To evaluate whether the difference of $H_{k}$ and $r$ between the two groups of networks results in the distinct performance of $k_S$ imprecision, we randomize the networks using two rewiring schemes (see \textbf{Methods} for details). In the first one, degrees of nodes are preserved after each single rewiring but correlations between the degrees of connected nodes are destroyed~\cite{maslov2002}. This keeps the $H_{k}$ unchanged with the original real networks while other structural features destroyed. As is shown in Fig.~S2 in SI, the coreness performance is greatly improved: the $k_S$ imprecision is very low and basically lower than or close to the $k$ imprecision in degree-preserving randomized networks. This indicates that the relatively small $H_{k}$ value of the second group of networks is not the reason of poor $k_S$ imprecision. Next, in the second scheme the rewiring preserves both the degrees of nodes and the joint degree-degree distribution of connected nodes, $P(k,k^{'})$, so that the degree-degree correlations of all nodes are preserved. This keeps both $H_{k}$ and $r$ unchanged as the original real networks, but as shown in Fig.~S3, the $k_S$ imprecision is very low and in general lower than the $k$ imprecision. This implies that the small $H_{k}$ value and positive $r$ are not the cause of poor $k_S$ imprecision in the second groups of networks. So, what is the real origin of the poor coreness performance?

\textbf{Analyzing the connectivity between shells.} We move to explore the complex connectivity between shells of the studied real networks. Specifically, we consider the link patterns from each shell to its upper shells (shells with greater $k_S$ index), equal shell (the shell with equal $k_S$ index) and lower shells (shells with smaller $k_S$ index). We define the link strength of node $i$ to its upper shells by the proportion function
\begin{equation}
r_{i}^{u}=\frac{e_{i}^{u}}{k_{i}},
\end{equation}
where $e_{i}^{u}$ is the number of links originating from node $i$ to nodes in upper shells , $k_{i}$ is the total number of links of node $i$, that is the degree of node $i$. Large $r_{i}^{u}$ indicates more links to the upper shells. Similarly, the link strengths of node $i$ to its equal shells and lower shells are quantified by
%\begin{equation}
$r_{i}^{e}=e_{i}^{e}/{k_{i}}$,
%\end{equation}
%\begin{equation}
$r_{i}^{l}=e_{i}^{l}/{k_{i}}$ respectively.
%\end{equation}
The link strengths of $k_S$-shell to its upper (equal, lower) shells are the average link strength of nodes in that shell, that is
\begin{equation}
R_{k_S}^{u(e, l)}=\frac{\sum_{i\in\Gamma_{k_S}}r_{i}^{u(e, l)}}{n_{k_S}}, %R_{k_S}^{e}=\frac{\sum_{i\in\Gamma_{k_S}}r_{i}^{e}}{n_{k_S}}, %R_{k_S}^{l}=\frac{\sum_{i\in\Gamma_{k_S}}r_{i}^{l}}{n_{k_S}},
\end{equation}
where $\Gamma_{k_S}$ consists of nodes with coreness $k_S$, $n_{k_S}$ is the number of nodes in $k_S$-shell, and $R_{k_S}^{u}+R_{k_S}^{e}+R_{k_S}^{l}=1$.

From Fig.~3 (a)-(c) for the first group of networks, $R^{u}_{k_S}$ generally decreases with $k_S$, this is because the number of nodes in upper shells decreases monotonously with the increase of $k_S$. $R^{e}_{k_S}$ remains stable with the increase of $k_S$. $R^{l}_{k_S}$ increases with $k_S$, and in the innermost core, this value goes up to $0.6$ and above. For large $k_S$, $R^{l}_{k_S}$ is much greater than $R^{e}_{k_S}$, which means a large proportion of links of high shells point to the their lower shells, obviously higher than the proportion of links within the shell. In Fig.~3 (d)-(f) for the second group, $R^{u}_{k_S}$ decreases with $k_S$ in Email and CA-Hep, although there is some fluctuation in Hamster. $R^{e}_{k_S}$ increases with $k_S$. {In the innermost core}, $R^{e}_{k_S}$ is close to $0.7$ in Email, close to $1.0$ in CA-Hep, and close to $0.8$ in Hamster, which is at least $50\%$ larger than that of the first group. $R^{l}_{k_S}$ increases with $k_S$ at first and falls suddenly at some high shells. For these three networks, $R^{l}_{k_S}$ are under $0.4$ in the innermost core, and is much lower than $R^{e}_{k_S}$. This indicates that in the second group, the proportion of links from high shells pointing to lower shells is obviously lower than the proportion of links pointing within the shell. This is a sign of densely connected small group within the shell. The average clustering coefficient of nodes in high shells also reflects the overly dense connection in high shells in the second group (See Fig.~S4 in SI). We plot the link strength of each shell to its lower shells, equal shell and upper shells in the degree-degree correlation preserving randomized networks in Fig. S5. $R^{l}_{k_S}$ is promoted above 0.35 and is greater than $R^{e}_{k_S}$ in most high shells in CA-Hep and Hamster, although in Email there is only a little promotion. {The rewiring has changed the dense local link patterns of core-like groups, which is reflect by the increase of $R^{l}_{k_S}$ and decrease of $R^{e}_{k_S}$ in high shells. The promoted $k_S$ performance in Fig. S3 is the result of enhanced link diversity.}

\begin{figure}[!ht]
\begin{center}
\epsfig{file=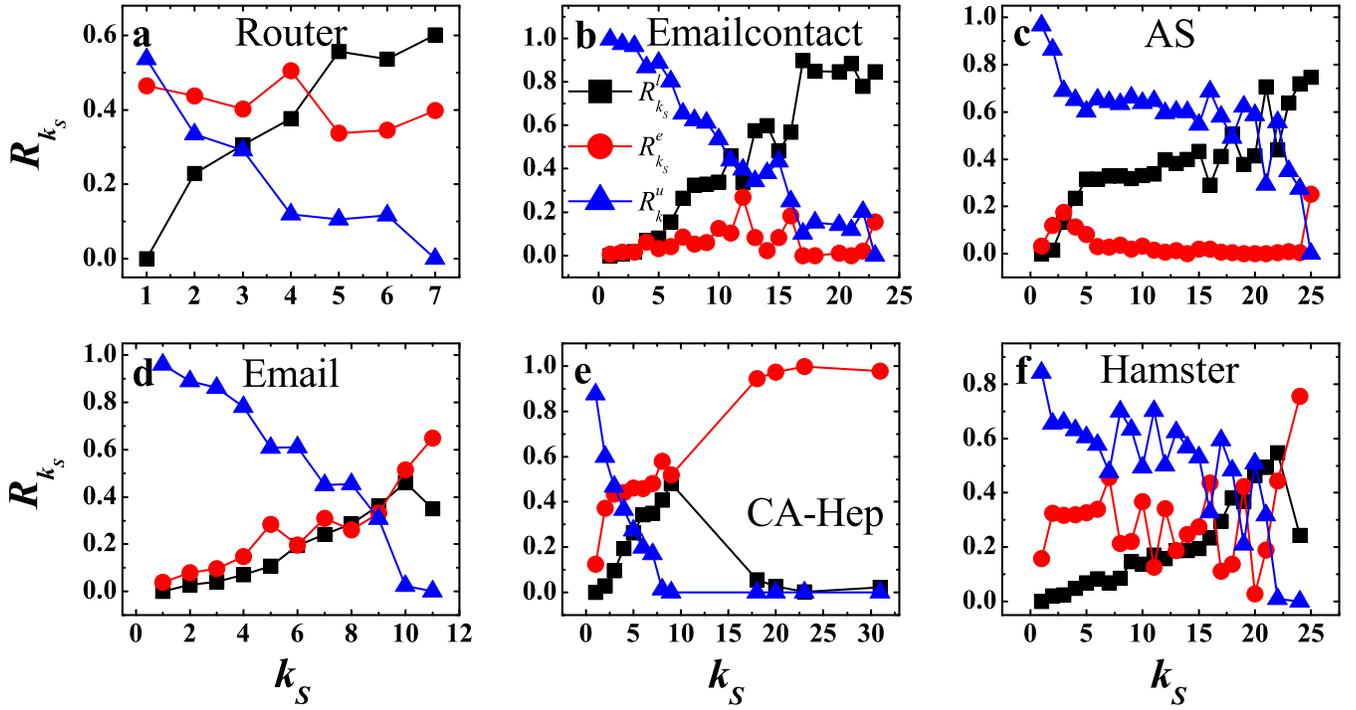,width=1\linewidth}

\caption{\textbf{Link strength of shells for the real networks.} The link strength of each shell to its lower shells $R^{l}_{k_S}$ (black squares), equal shell $R^{e}_{k_S}$ (red circles) and upper shells $R^{u}_{k_S}$ (blue triangles) are represented. $k_S$ ranges from the smallest $k_S$ value to the largest $k_S$ value in the network.}
\label{figure3}
\end{center}
\end{figure}

Next, we focus on the link pattern of the innermost core. The link strength $r^{ks}_{i}=e^{ks}_{i}/k_{i}$ defines the ratio of links from a innermost core node $i$ to the shell with index $k_S$ to the degree of node $i$. $R_{k_{Smax}}$ is the average link strength of nodes in the innermost core to the shell with index $k_S$. Fig.~4 (a) shows the link strength of innermost core to all shells in the first three networks, which is a U-shape curve. In these networks, apart from the link ratio within the core, the largest link ratio points to the shell with most nodes, usually the 1-shell. A U-shape distribution of links from the core is a good feature of core-periphery structure, in which core nodes are well connected to other core nodes and to periphery nodes and periphery nodes are not well connected to each other~\cite{puck2012}. In the second group, shown in Fig.~4 (b), the link of innermost core to all shells is different from the first group. Core nodes are very inclined to connecting to core nodes, with a link strength above $0.6$. The second largest link ratio points to the adjacent shell of the innermost core, other than the shell with most nodes. When an epidemic spreading origins from core nodes, it is easy to spread throughout the core, but is relatively difficult to spread system wide. This locally connected phenomenon also implies the origin of core-like group (i. e., false core): nodes are densely connected within a small group which contributes much to the $k_S$ index of the nodes, but in the whole network these nodes are not best connected and not located in the most important position for spreading. The link pattern of the second innermost shell is shown in Fig.~S6 in SI.

\begin{figure}[!ht]
\begin{center}
\epsfig{file=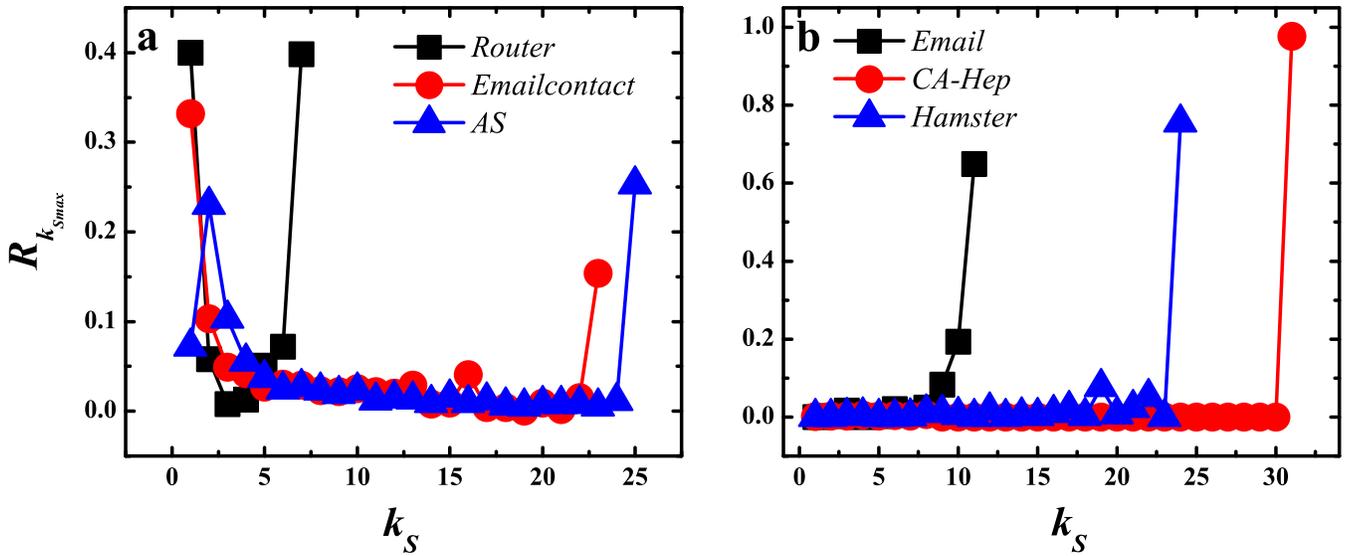,width=1\linewidth}

\caption{\textbf{Link strength of the innermost core to each shell of the network. }(a) The link strength of the innermost core to each shell exhibits a U-shape curve in Router (black squares), Emailcontact (red circles) and AS (blue triangles) networks. (b) The link strength of the innermost core to each shell exhibit a slope in Email (black squares), CA-Hep (red circles) and Hamster (blue triangles) networks. $k_S$ ranges from the smallest $k_S$ value to the largest $k_S$ value in the network.}
\label{figure4}
\end{center}
\end{figure}

\begin{figure}[!ht]
\begin{center}
\epsfig{file=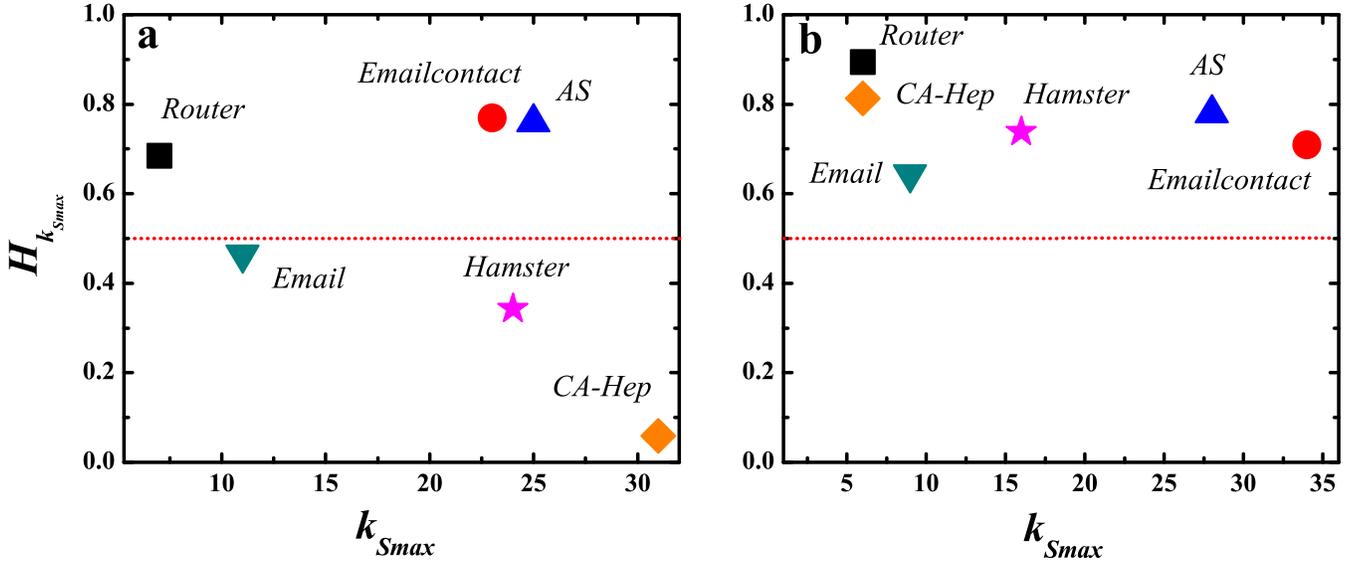,width=1\linewidth}
\caption{\textbf{Link entropy of the innermost core for the real networks and their randomized version.} (a) Link entropy of the innermost core for the real networks.  (b) Link entropy of the innermost core for the degree-preserving randomized networks.  $k_{Smax}$ is the largest $k_S$ value in the network. $H_{k_{Smax}}$ is the link entropy of the innermost core.}
\label{figure5}
\end{center}
\end{figure}

\textbf{Identifying core-like groups from a structural perspective.} The above analysis suggests an obvious structural difference between the two groups of networks: in the first one, the link pattern of innermost core to other shells exhibits a strong diversity, while in the second group, the link of innermost core is very localized within the shell. To quantify the link diversity of a shell with index $k_S$, we define a link entropy as
\begin{equation}\label{entropy}
H_{k_S}=-\frac{1}{lnL}\sum_{k'_S=1}^{k_{Smax}}r_{k_S,k'_S}lnr_{k_S,k'_S},
\end{equation}
where $r_{k_S,k'_S}$ is the average link strength of $k_S$-shell to the $k'_S$-shell and $L$ is the number of shells. The normalized factor $lnL$ measures the entropy when links are uniformly distributed in all shells. This normalization makes the networks with different number of shells comparable. For the innermost core of each network, $k_S$ is set to the maximum $k_S$ value of the network. Entropy of cores of the real network and its degree-preserving randomized version are shown in Fig.~\ref{figure5}. In Fig.~\ref{figure5} (a), true cores have a link entropy $H_{k_{Smax}}$ higher than $0.6$ while false cores have a link entropy lower than $0.5$. But in the randomized network, Fig.~\ref{figure5} (b), all the cores have a link entropy $H_{k_{Smax}}$ higher than $0.6$. Fig. S7 in SI shows the core entropy of degree-degree preserving randomized networks, which is above 0.5 for all studied networks. High entropy corresponds to a more uniform link pattern, where the core is well-connected to the other parts of the network. Low entropy corresponds to a localized link pattern, where the core is densely connected within the shell. In fact, these false cores are not located in the central position of the networks, reflecting by the relatively low spreading efficiency, e.g. the 11-shell in Email, 31-shell in CA-Hep and 24-shell in Hamster, as shown in Figs. S9 and S10 in SI.

\textbf{Locating the position of core-like groups throughout the networks.} Uncovering locally connected core-like groups leads us to understand the imprecision of coreness centrality in the spreading process. We present the imprecision function of PGP, Netsci and Astro networks in Fig.~\ref{figure6}(a)-(c). The coreness performs very well at large $k_S$ values, but then rises suddenly at certain shells. Specifically speaking, in PGP at the 22-shell and above, the imprecision of $k_S$ is lower than that of $k$. However, there are sudden rises at the 21-shell, 16-shell and 15-shell. From the 10-shell, the $k_S$ imprecision is lower than $k$ again. In Netsci, the $k_S$ imprecision is very low at the 8-shell. Then it rises up at the 7-shell and 6-shell. The $k_S$ imprecision is worse than that of $k$ until the 4-shell. In Astro, the $k_S$ imprecision is low at the 51-shell and higher shells. Then it rises up at the 48-shell and then falls. The same phenomenon occurs at the 30-shell. The rise of $k_S$ imprecision implies that the corresponding shells are core-like groups. Locating them by a dynamic spreading method requires time-consuming simulations.

According to Eq.~(\ref{entropy}), we calculate the link entropy of each shell in these networks. The shells outlined by hollow red circles in Fig.~\ref{figure6} (d)-(f) have relatively low entropy, which corresponds to locally connected core-like groups. This is reflected by the rise in $k_S$ imprecision shown in Fig.~\ref{figure6} (a)-(c). The link patterns of the core-like groups, shown in Fig. S8, are similar to that of false cores in the second group of Email, CA-Hep and Hamster: a dense connection within the shell. The only difference is that these core-like groups locate in the outer shells of the network other than the innermost shell. These core-like groups have an obvious low spreading efficiency than their adjacent shells, which is also confirmed in Fig. S9 and S10. From the above, we see the link entropy provides a fast way to locate the position of core-like groups in the network without running a large amount of spreading simulations, which is very important in identifying key spreaders and controlling the spreading dynamics on networks.

\begin{figure}[!ht]
\begin{center}
\epsfig{file=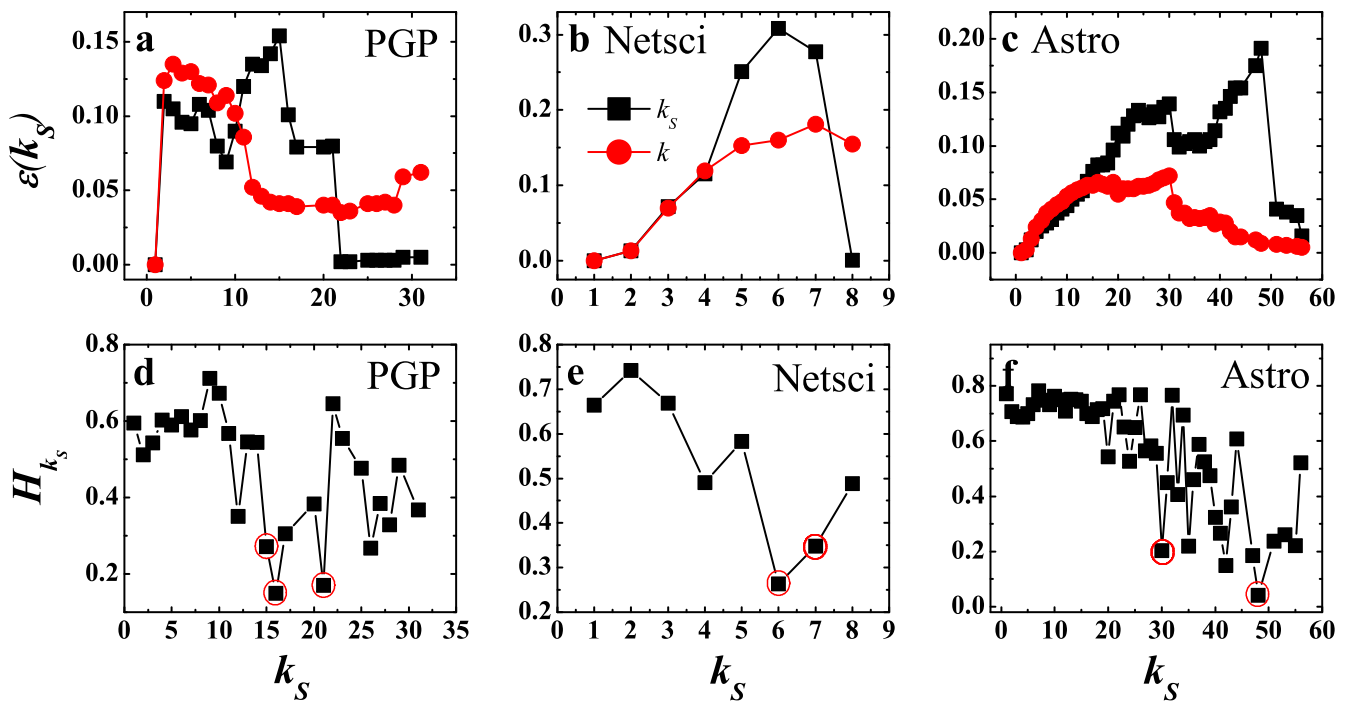,width=1\linewidth}
\caption{\textbf{Locating core-like groups in real networks by link entropy.} (a)-(c) The imprecision of $k_S$ and $k$ as a function of $k_S$ for three real networks. The $k_S$ imprecision (black squares) and $k$ imprecision (red circles) are compared. Link entropy of shells in three networks. $H_{k_S}$ is the link entropy of $k_S$ shell. Hollow red circles outline the shells which are densely connected core-like groups. These are 21-shell, 16-shell and 15-shell in PGP, 7-shell and 6-shell in Netsci and 48-shell and 30-shell in Astro. $k_S$ ranges from the smallest $k_S$ value to the largest $k_S$ value in the network. }
\label{figure6}
\end{center}
\end{figure}

\section*{Discussion}

Analyzing and profiling the structures of real networks is an important step in understanding and controlling dynamic behaviors on networks. The \emph{k}-shell decomposition is a powerful tool to profile the hierarchical structures of networks. The inner core corresponds to the shells of large $k_S$ and the network periphery corresponds to the shells of small $k_S$. This  makes $k_S$ index an effective centrality measure to distinguish the spreading capability of nodes, which is validated in many real networks. However, there are circumstances where the \emph{k}-shell decomposition is not able to identify influential spreaders, which leaves much space to explore. Here from the perspective of core's spreading efficiency, we discover that in some real networks, there exist core-like groups, which have high coreness but are in fact not located in the core of the network. By analyzing the \emph{k}-core structure of real networks, we discover the distinct link patterns of true cores and core-like groups. For the true core of a network, it displays strong link diversity to other shells of the network, represented by a U-shape link curve. As for the core-like group, it has a very dense and local internal connection, represented as a slope-shape link curve. Based on the link pattern, we define a measure of link entropy to evaluate the link diversity of a shell to the remaining shells of the network. This provides a fast way to locate the core-like groups throughout the network from a structural perspective, which have a relatively low link entropy. We note that in Ref. [29] the authors calculated the entropy of each node to assess the heterogeneity of links. They use \emph{k}-shell decomposition to assign each node a global feature and compute for each node an entropy as its global diversity, which is then combined with local feature to rank node influence. However, this entropy relates much to the degree of a node. A node with small degree has a limited number of layers it can connect to, even the connection is uniformly distributed. The node entropy is limited in the sense of statistics. Contrary to their work, we target to a group. We consider the link diversity of a shell, which consists of several nodes and these nodes have different degrees. By using the entropy for a shell, we can effectively locate the core-like groups, whose \emph{k}-shell index is unable to reflect their global importance. This makes implication to the works that use the \emph{k}-shell index in ranking node importance.

Uncovering these core-like groups is important in identifying key players and making control strategy for spreading dynamics. It is worth noticing that in the core-like groups, there may also exist some good spreaders. It implies that there should be new network analysis method which will effectively locate the nodes of different importance in core-like groups in the right hierarchical position. The new method should apply well in real networks with specific structures such as strong community structures.

\section*{Methods}
\textbf{\emph{The $k$-shell decomposition}.}
The algorithm starts by removing all nodes with degree $k=1$. After removing all nodes with $k=1$, there may appear some nodes with only one link left. We should iteratively remove these nodes until there is no node left with $k=1$. The removed nodes are assigned with an index $k_S=1$ and are considered in the 1-shell. In a similar way, nodes with degree $k=2$ are iteratively removed and assigned an index $k_S=2$. This pruning process continues removing higher shells until all nodes are removed. As a result, each node is assigned a $k_S$ index, and the network can be viewed as a hierarchical structure from the innermost shell to the periphery shell.

\textbf{\emph{SIR Model}}.
The Susceptible-Infected-Recovered (SIR) model is widely used for simulating the spreading process on networks. In the model, a node has three possible states: $S$ (susceptible), $I$ (infected) and $R$ (recovered). An individual in the susceptible state does not have the disease yet but could catch it if they come into contact with someone who does. An individual in the infected state has the disease and can pass it to susceptible individuals. An individual in recovered state neither spread disease nor be infected by others. In the start of a spreading process, a single node is infected, considered as seed, and all other nodes are in susceptible states. At each time step, there are two stages. In the first stage, susceptible individuals become infected with probability $\lambda$ when they have contacted with an infected neighbor. In the second stage, infected nodes recover or die (change to $R$ state) with probability $\mu$. Here we set $\mu=1$ for generality. The spreading process stops when there is no infected node in the network. The proportion of recovered nodes defines the final infection population in a spreading process. We record the average infected population $M_{i}$ originating at node $i$ over $100$ times of the spreading process to quantify the influence of node $i$ in a SIR spreading.

As we take the final infected population to quantify the spreading efficiency of each node, the infection probability should be carefully considered. If it is too large, the effect of node position is not obvious and all nodes show almost identical spreading capabilities. If it is too small, the infection is very localized in the neighborhood, which cannot reflect the overall spreading influence of the nodes. So we first calculate the epidemic threshold of a network using the heterogeneous mean-filed method in Ref.~\cite{castellano2010}. That is $\lambda_{c}=\langle k\rangle/(\langle k^{2}\rangle-\langle k\rangle)$. Then we chose an infection probability $\lambda>\lambda_{c}$~\cite{macdonald2012,bae2014}, which makes the final infected population above the critical point, $M>0$, and reaches a finite but small fraction of the network size for most nodes as spreading origins, in the range of $1\%$-$20\%$~\cite{kitsak2010}. In fact, we plot the infected population of a shell as an average over nodes belong to the shell when infection probability is $1$-$5$ times of the threshold $\lambda_{c}$, as well as the infected population when infection probability is around the chosen infected probability $\lambda$. We find that, the relative spreading efficiency of shells is almost the same under different infection probabilities (See Fig. S9 and S10 in SI).

\textbf{\emph{Rewiring Schemes}.}
In the first rewiring scheme, we randomly choose two edges of the network, and label the ends of the first edge as A and B, and the ends of the second edge as C and D. Then we rewire the two edges, connecting end A and D as an edge, and connecting end B and C as another edge. We avoid multiple edge and self-edge in the rewiring process. This rewiring preserves the degree sequence of the original real network but destroys the degree correlations. In the second rewiring scheme, we randomly choose an edge and test the degree of one end, record as $k$. A second edge with an end having degree $k$ is then chosen. We rewire the two edges as before and ensure that the end connecting to a node of degree $k$ still connects to a node of degree $k$ after rewiring. This scheme preserves both the degree sequence and the degree-degree correlations as the original real network.

\textbf{\emph{Data Sets}.}
The real networks studied in the paper are: (1) Router (the router level topology of the Internet, collected by the Rocketfuel Project)~\cite{spring2004};(2) Email-contact (Email contacts at Computer Science Department of University college London)~\cite{kitsak2010}; (3) AS (Internet at the autonomous system level)~\cite{newmandataas}; (4) Email (e-mail network of University at Rovira i Virgili, URV) ~\cite{guimera2003};(5) CA-Hep (Giant connected component of collaboration network of arxiv in high-energy physics theory)~\cite{leskovec2012}; (6) Hamster (friendships and family links between users of the website hamsterster.com)~\cite{hamster2014}; (7) PGP (an encrypted communication network)~\cite{boguna2004}; (8) Netsci (collaboration network of network scientists)~\cite{newman2006}; (9) Astro physics (collaboration network of astrophysics scientists)~\cite{newman2001}.

\section*{Figure legends}

{\bf Figure 1}: The imprecision of $k_S$ and $k$ as a function of $p$ for nine real networks. The $k_S$ imprecision (black squares) and $k$ imprecision (red circles) are compared in each network. $p$ is the proportion of nodes calculated, ranging from 0.003 to 0.029.  See Fig.S1 for large $p$ plots in SI.

{\bf Figure 2}:
The imprecision of $k_S$ and $k$ as a function of $k_S$ for six real networks. The $k_S$ imprecision (black squares) and $k$ imprecision (red circles) are compared in each network. Each square represents the $k_S$ imprecision of nodes in $k_S$-core, and each circle represents the $k$ imprecision of $n$ highest degree nodes, where $n$ equals to the number of nodes in $k_S$-core. $k_S$ is an integer representing the shell index, ranging from the smallest $k_S$ value to the largest $k_S$ value in the network.

{\bf Figure 3}:
Link strength of shells for the real networks. The link strength of each shell to its lower shells $R^{l}_{k_S}$ (black squares), equal shell $R^{e}_{k_S}$ (red circles) and upper shells $R^{u}_{k_S}$ (blue triangles) are represented. $k_S$ ranges from the smallest $k_S$ value to the largest $k_S$ value in the network.

{\bf Figure 4}: Link strength of the innermost core to each shell of the network. (a) The link strength of the innermost core to each shell exhibits a U-shape curve in Router (black squares), Emailcontact (red circles) and AS (blue triangles) networks. (b) The link strength of the innermost core to each shell exhibit a slope in Email (black squares), CA-Hep (red circles) and Hamster (blue triangles) networks. $k_S$ ranges from the smallest $k_S$ value to the largest $k_S$ value in the network.

{\bf Figure 5}: Link entropy of the innermost core for the real networks and their randomized version. (a) Link entropy of the innermost core for the real networks.  (b) Link entropy of the innermost core for the degree-preserving randomized networks.  $k_{Smax}$ is the largest $k_S$ value in the network. $H_{k_{Smax}}$ is the link entropy of the innermost core.

{\bf Figure 6}: Locating core-like groups in real networks by link entropy. (a)-(c) The imprecision of $k_S$ and $k$ as a function of $k_S$ for three real networks. The $k_S$ imprecision(black squares) and $k$ imprecision(red circles) are compared. Link entropy of shells in three networks. $H_{k_S}$ is the link entropy of $k_S$ shell. Hollow red circles outline the shells which are densely connected core-like groups. These are 21-shell, 16-shell and 15-shell in PGP, 7-shell and 6-shell in Netsci and 48-shell and 30-shell in Astro. $k_S$ ranges from the smallest $k_S$ value to the largest $k_S$ value in the network. \\

\section*{Acknowledgement}
This work was partially supported by the National Natural Science Foundation of
China (Grant Nos. 11105025, 91324002, 61433014), Scientific Research Starting Project of Southwest Petroleum University (No. 2014QHZ024) and Youth Foundation of Southwest
Petroleum University (Grant No. 285).
Y. Do was supported by Basic Science Research Program through
the National Research Foundation of Korea (NRF) funded by the Ministry of Education,
Science and Technology (NRF-2013R1A1A2010067).

\section*{Author contributions}
Y. L., M. T. and T. Z. devised the research project.
Y. L. performed numerical simulations.
Y. L., M. T. and Y. H. D. analyzed the results.
Y. L., M. T., T. Z. and Y. H. D. wrote the paper.

\section*{Additional information}

%{\bf Supporting Information} accompanies this paper at

%http://www.nature.com/scientificreports

{\bf Competing financial interests}:
The authors declare no competing financial interests.

~\\
\begin{center}
{\Large
Supporting Information for\\
\vspace{0.5cm}
\textbf{Core-like groups result in invalidation of identifying super-spreader by k-shell decomposition}
}\\
\vspace{0.5cm}

\large{Ying Liu, Ming Tang, Tao Zhou and Younghae Do}

\end{center}

\renewcommand\thetable{S\arabic{table}}
\begin{table}[!ht]
\setcounter{table}{0}
\caption{\textbf{Proportion of nodes in high shells in the studied real networks.} $d$ indicates the shell ranking difference of a shell from the highest shell. $d=0$ corresponds to the highest shell, $d=1$ corresponds to the nearest shell from the highest shell. For example, the highest shell of CA-Hep is 31-shell, while the second highest shell is 23-shell. Thus, $d=0$ corresponds to 31-shell, and $d=1$ correspond to 23-shell.}\centering
\begin{tabular}{ccccccccccc}
\hline
\hline
\textbf{Network} & \textbf{$d=0$} & \textbf{$d=1$} & \textbf{$d=2$} & \textbf{$d=3$} & \textbf{$d=4$} & \textbf{$d=5$} & \textbf{$d=6$} & \textbf{$d=7$}& \textbf{$d=8$} & \textbf{$d=9$} \\
\hline
%\rowcolor{mygray}
Router &0.52\% &0.80\% &1.25\% &2.15\% &3.25\% &6.53\% &85.50\% &0 &0 &0\\
%\hline
%\rowcolor{mygray}
Emailcontact &0.33\%	&0.05\% &0.01\%	&0.03\%	&0.02\%	&0.02\%	&0.29\%	&0.10\%	&0.06\%	&0.26\%\\
%\hline
%\rowcolor{mygray}
AS &0.31\% &0.03\%	&0.02\%	&0.03\%	&0.03\%	&0.04\%	&0.03\%	&0.03\%	&0.06\%	&0.06\%\\
\hline
%\rowcolor{mypink}
Email &1.06\%	&9.62\%	&10.33\%	&9.80\%	&7.50\%	&8.83\%	&11.56\%	&7.33\%	&8.83\%	&11.47\%\\
%\hline
%\rowcolor{mypink}
CA-Hep &0.37\%	&0.28\%	&0.24\%	&0.22\%	&0.12\%	&2.07\%	&3.62\%	&6.44\%	&10.23\%	&13.36\%\\
%\hline
%\rowcolor{mypink}
Hamster &1.25\%	&4.25\%	&2.75\%	&0.70\%	&2.75\%	&1.85\%	&1.30\%	&6.25\%	&2.90\%	&3.40\%\\
\hline
%\rowcolor{mycyan}
PGP &0.38\%	&0.02\%	&0.67\%	&0.06\%	&0.03\%	&0.02\%	&0.19\%	&0.02\%	&0.23\%	&0.01\%\\
%\hline
%\rowcolor{mycyan}
Netsci &2.38\% &4.22\% &5.54\%	&6.07\%	&26.91\% &24.80\% &22.96\%	&7.12\%	&0	&0\\
%\hline
%\rowcolor{mycyan}
Astro &0.38\% &0.38\% &0.35\%	&0.38\% &0.33\% &0.30\% &0.28\%	&0.01\%	&0.29\%	&0.56\%\\
\hline
\hline
\end{tabular}
%\begin{flushleft}
%\end{flushleft}
\label{tab:proportion}
\end{table}

\begin{figure}[!ht]
\begin{center}
\epsfig{file=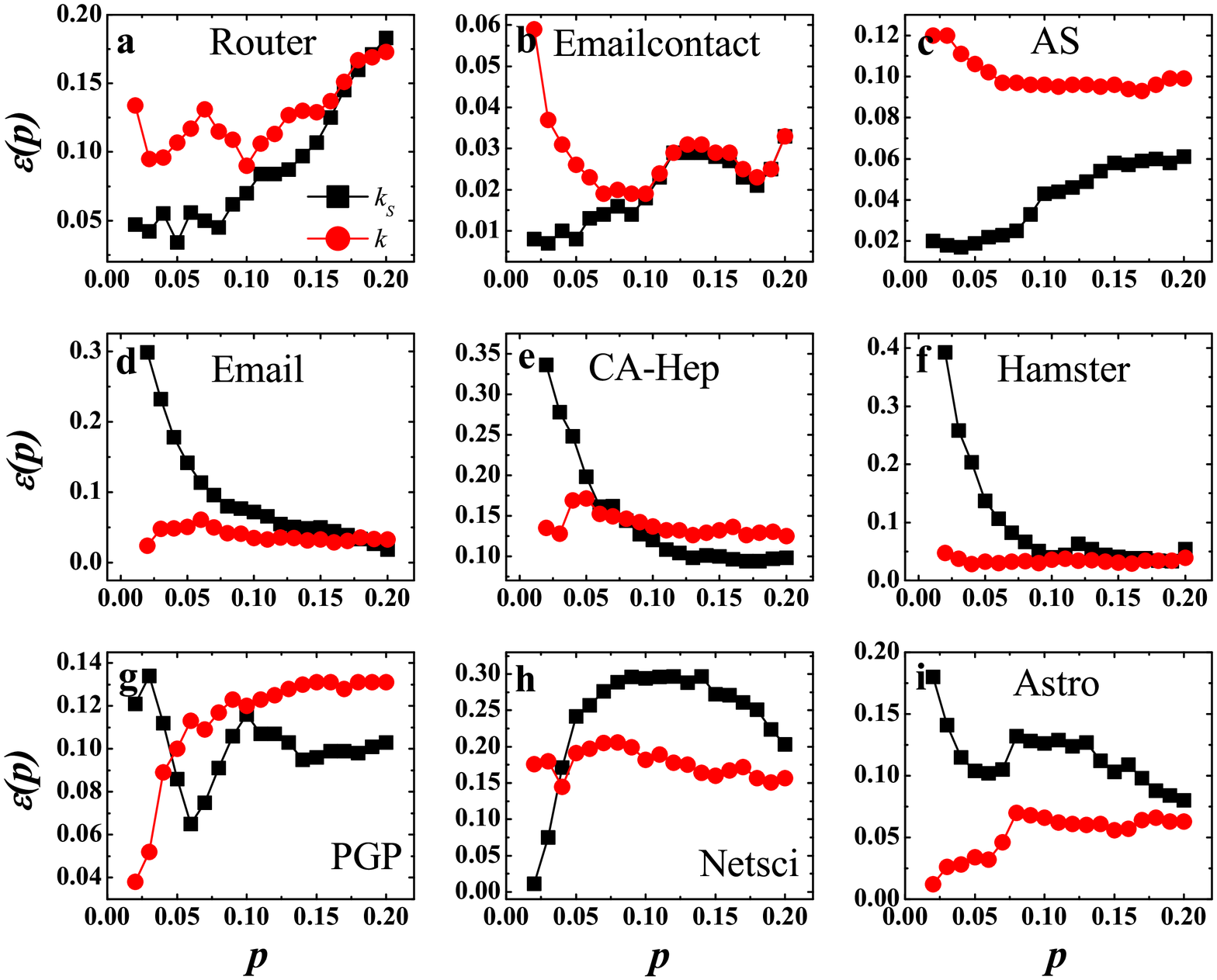,width=1\linewidth}
\setcounter{figure}{0}
\renewcommand\thefigure{S\arabic{figure}}
\caption{\textbf{The imprecision of $k_S$ and $k$ as a function of $p$ for nine real networks.} $p$ is the proportion of nodes calculated, ranging from 0.02 to 0.2. }
\label{figures1}
\end{center}
\end{figure}

\textbf{Text S1. Explanation of the $k_S$ imprecision demonstrated in Fig. S1.}
 In Router, the imprecision is under 0.1 at first, and then rises at around $p\approx0.15$. As the number of nodes in 1-shell accounts for $85.5\%$ of the network size, the rising results from the random selection of nodes in 1-shell. In Emailcontact and AS, within in $20\%$ of the network size, the imprecision is low. In Fig. S1 (d)-(f) the imprecision of $k_S$ decreases with $p$. In Email, the core-like groups, 11-shell, accounts for $1.1\%$ of the network size. From 10-shell, the spreading efficiency decreases with shells. The number of nodes in 10-shell accounts for about $10\%$ of the network size. This results in the sharp decrease of $k_S$ impression until $p\approx 0.11$. In CA-Hep, the 31-shell, 23-shell, 20-shell and 18-shell accounts for about $1\%$ of network size, which corresponds to the high $k_S$ imprecision for $p<0.01$ (See Fig. 1 in main text). Next to the 18-shell is the 9-shell. From the 9-shell, the spreading efficiency decreases with shells. The proportion of nodes in shells 9, 8, 7, 6 is $0.1\%$, $2\%$, $3.6\%$, $6.4\%$ respectively. Thus at $p\approx0.13$, the imprecision of $k_S$ goes to a low value. In Hamster, the innermost 24-shell accounts for $1.25\%$ of the network size, which corresponds to the high $k_S$ imprecision before $p=0.0125$ (See Fig. 1 in main text). Then the $k_S$ imprecision decreases with $p$ until $p\approx 0.09$, which corresponds to the high spreading efficiency of 22-shell and 21-shell. In PGP, from the 31-shell to 22-shell, accounting for $1.4\%$ of network size, the spreading efficiency decrease from the inner shell to periphery shell, which corresponds to the low $k_S$ imprecision (See Fig. 1 in main text). Then from 21-shell to 11-shell, locally densely connected groups occurs. The proportion of nodes of these shells accounts for $2.1\%$ of the network size. Thus , at $p\approx 0.035$, the $k_S$ imprecision goes down. In Netsci, core nodes accounts for $2.4\%$ of network size, that corresponds to the the low $k_S$ imprecision before $p=0.024$.  Then in shell 7 and 6 , where the proportion of nodes are $4.2\%$, $5.5\%$ respectively, locally densely connected groups occurs, which corresponds to the highest $k_S$ imprecision at $p\approx 0.12$. The proportion of nodes in 5-shell is $6.1\%$, and since then, the $k_S$ imprecision begins to decrease. In Astro, the low $k_S$ imprecision before $p=0.015$ (See Fig. 1 in main text) corresponds to the 51-shell and above. The 48-shell is a local group, accounting for $0.3\%$ of the the network size. This corresponds to the sharp rise at $p\approx0.015$. Then the $k_S$ imprecision goes down.

\begin{figure}[!ht]
\begin{center}
\epsfig{file=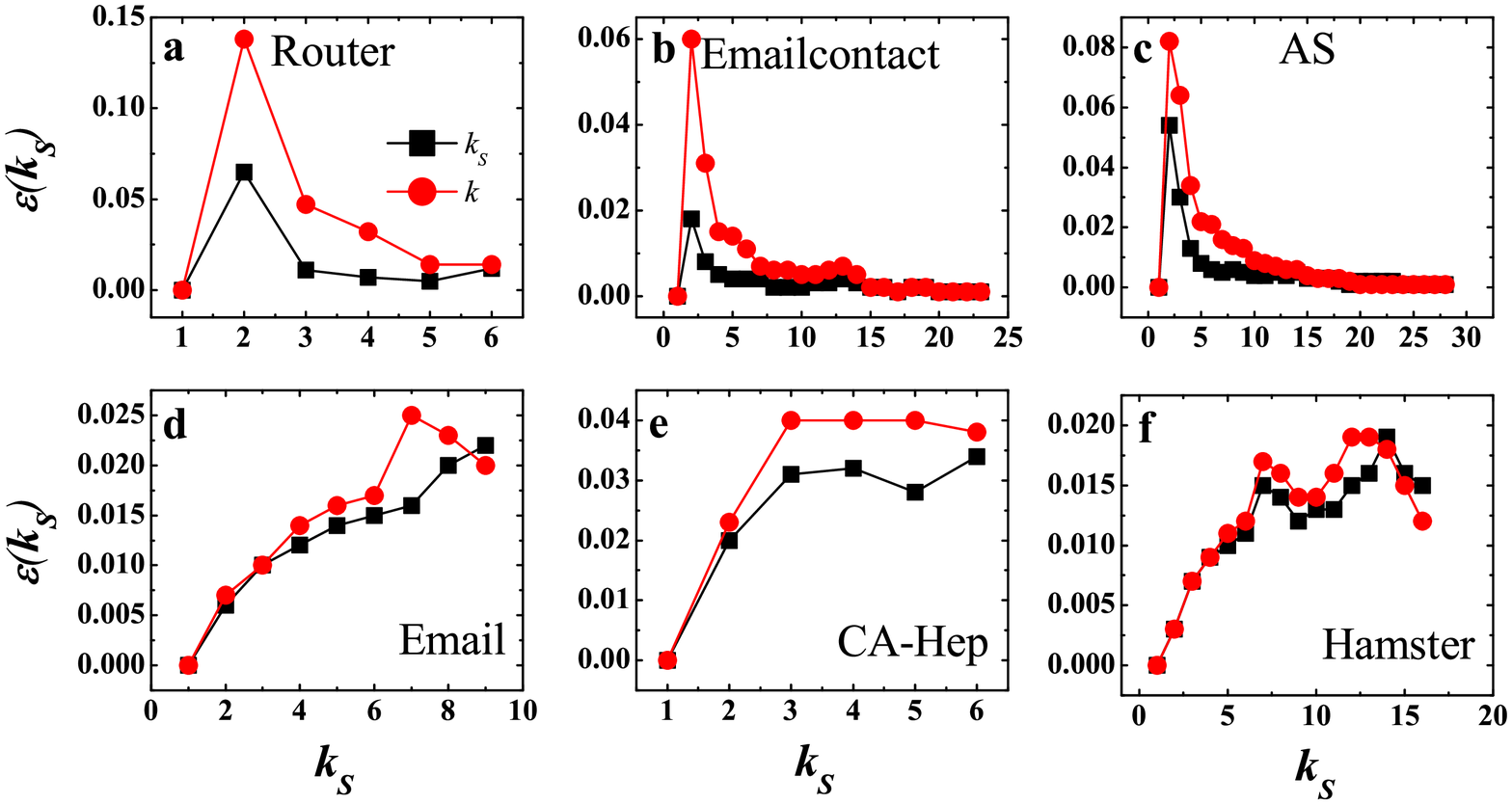,width=1\linewidth}
\renewcommand\thefigure{S\arabic{figure}}
\caption{\textbf{The imprecision of $k_S$ and $k$ as a function of $k_S$ for degree-preserving randomized networks.} In all the networks shown in (a)-(f), the $k_S$ imprecision is very low, under the value of 0.07, and in most cases lower than $k$ imprecision. For the randomized networks of Email and Hamster, although the $k_S$ imprecision is slightly higher than that of degree in some shells, the absolute values are very low, under 0.025. This indicates that the $k$-shell strategy is more effective than or at least as well as degree in most cases.}
\label{figures2}
\end{center}
\end{figure}

\begin{figure}[!ht]
\begin{center}
\epsfig{file=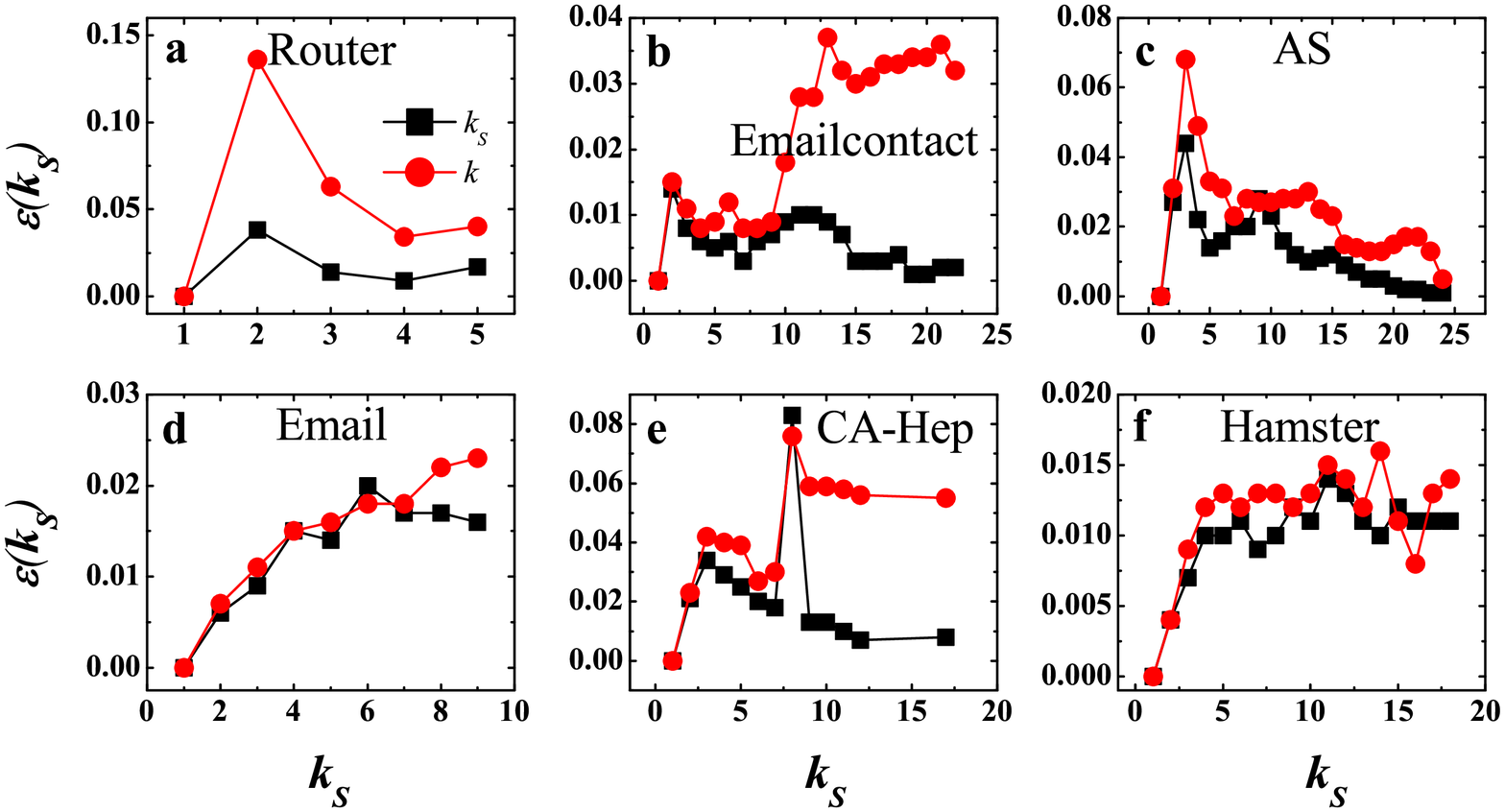,width=1\linewidth}
\renewcommand\thefigure{S\arabic{figure}}
\caption{\textbf{The imprecision of $k_S$ and $k$ as a function of $k_S$ for degree-degree correlation preserving randomized networks. }The imprecision of $k_S$ is very low in high shells and is lower than that of $k$. }
\label{figures3}
\end{center}
\end{figure}

\begin{figure}[!ht]
\begin{center}
\epsfig{file=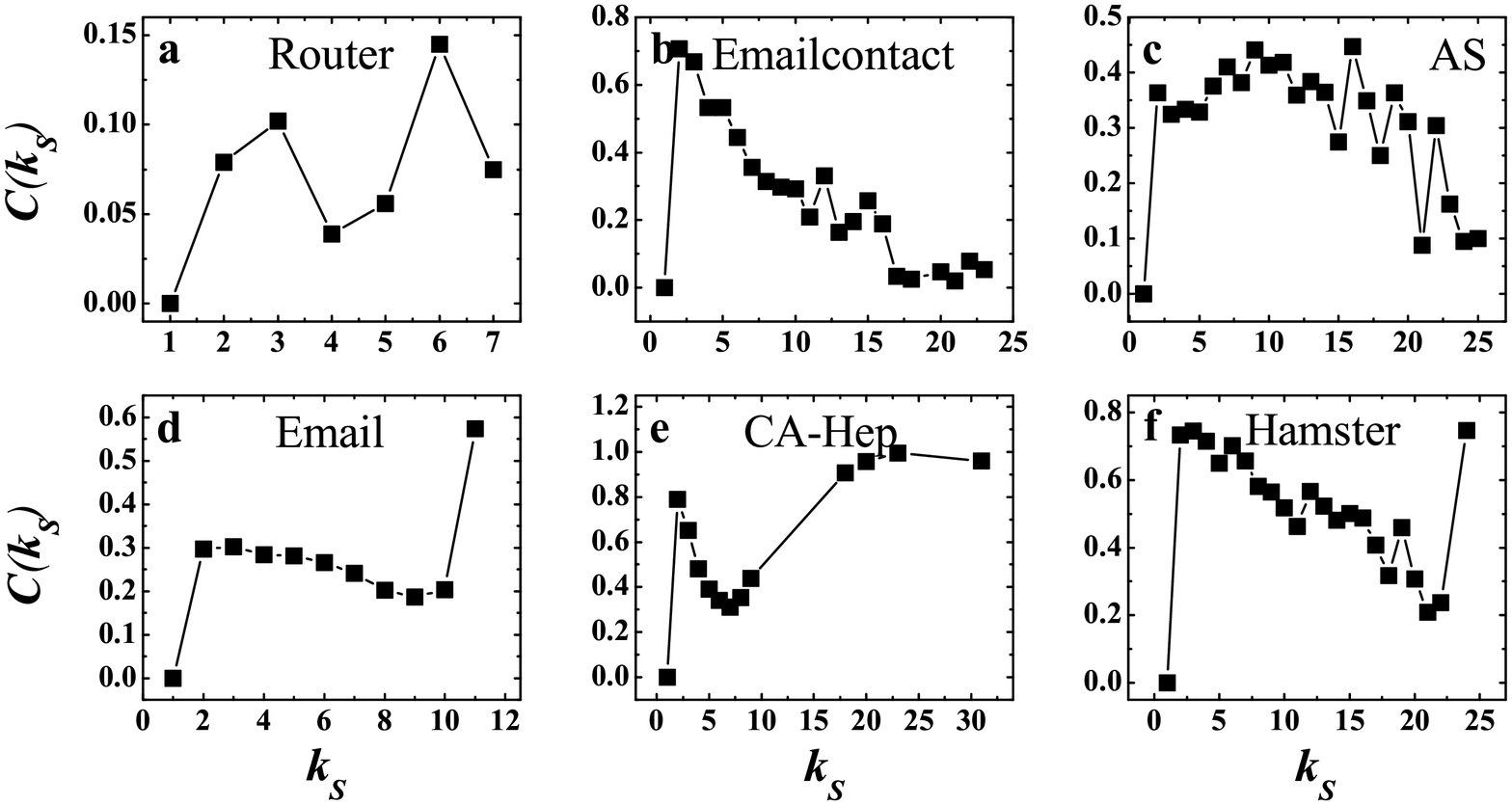,width=1\linewidth}
\renewcommand\thefigure{S\arabic{figure}}
\caption{\textbf{Clustering coefficient of shells for the real networks.} (a), (b), (c) The average clustering coefficient is smaller than 0.1 in the innermost core shell. (d), (e), (f) The average clustering coefficient is greater than 0.5 in the innermost core. }
\label{figures4}
\end{center}
\end{figure}

\begin{figure}[!ht]
\begin{center}
\epsfig{file=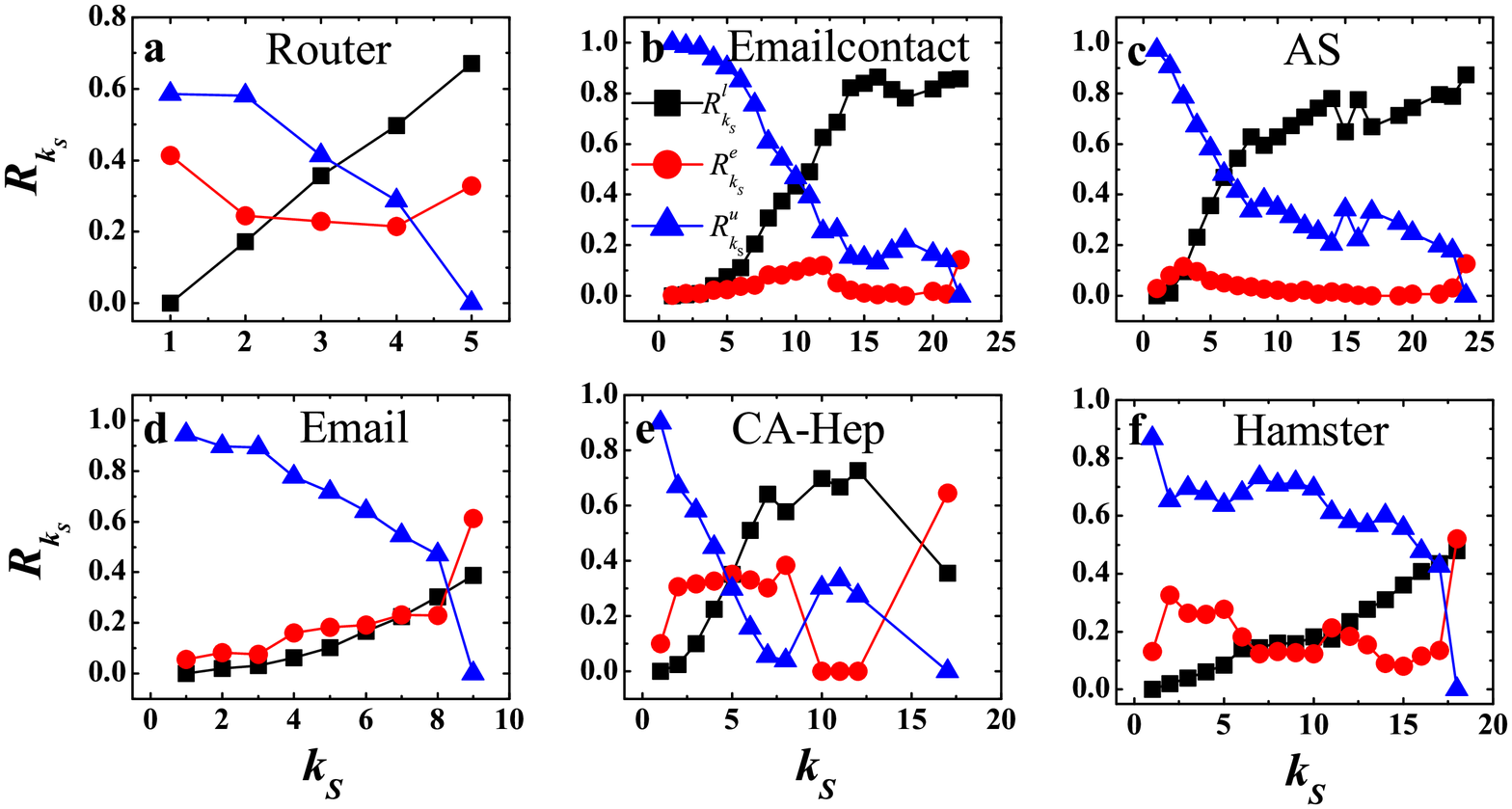,width=1\linewidth}
\renewcommand\thefigure{S\arabic{figure}}
\caption{\textbf{Link strength of shells for degree-degree correlation preserving randomized networks .} The link strength of each shell to its lower shells $R^{l}_{k_S}$ (black squares), equal shell $R^{e}_{k_S}$ (red circles), and upper shells $R^{u}_{k_S}$ (blue triangles) in the degree-degree correlation preserving randomized networks are represented. (a), (b), (c) $R^{l}_{k_S}$ is much larger than $R^{e}_{k_S}$ in high shells. (d), (e), (f) $R^{l}_{k_S}$ is strongly promoted and is always larger than $R^{e}_{k_S}$ in high shells in CA-Hep (e) and Hamster (f), although in Email (d) there is no obvious promotion. }
\label{figures5}
\end{center}
\end{figure}

\begin{figure}[!ht]
\begin{center}
\epsfig{file=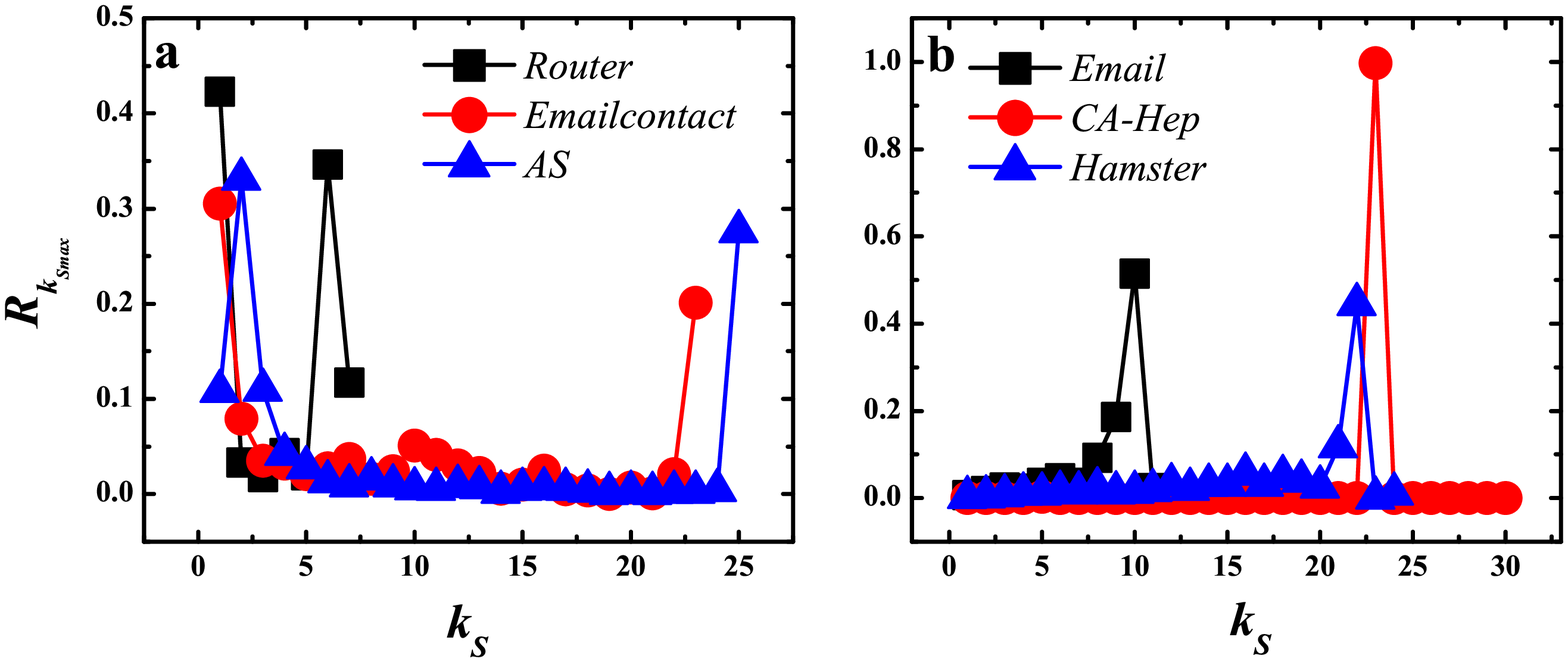,width=1\linewidth}
\renewcommand\thefigure{S\arabic{figure}}
\caption{\textbf{Link strength of the second innermost shell to each shell of the network.} (a) Similar to the core, the second innermost shell are well connected to other parts of the network in Router (black squares), Emailcontact (red circles) and AS (blue triangles). (b) The link ratio within the second highest shell is lower than $0.6$ in Email (black squares) and Hamster (blue triangles), but is still close to $1.0$ CA-Hep (red circles).}
\label{figures6}
\end{center}
\end{figure}

\begin{figure}[!ht]
\begin{center}
\epsfig{file=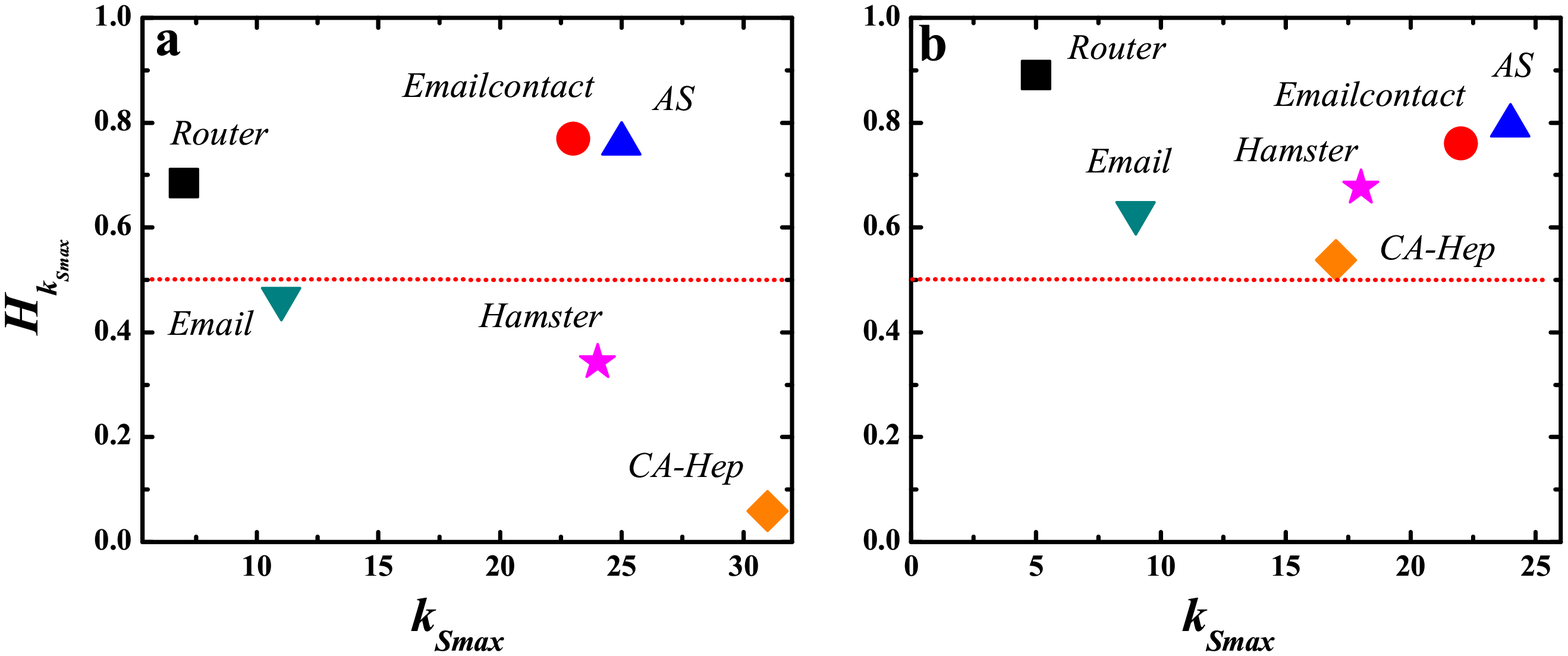,width=1\linewidth}
\renewcommand\thefigure{S\arabic{figure}}
\caption{\textbf{Link entropy of the innermost core for the real networks and their randomized version.} (a) Link entropy of the innermost core for the real networks. (b) Link entropy of the innermost core for the degree-degree correlation preserving randomized networks. In all the randomized networks, the core entropy is above 0.5.}
\label{figures7}
\end{center}
\end{figure}

\begin{figure}[!ht]
\begin{center}
\epsfig{file=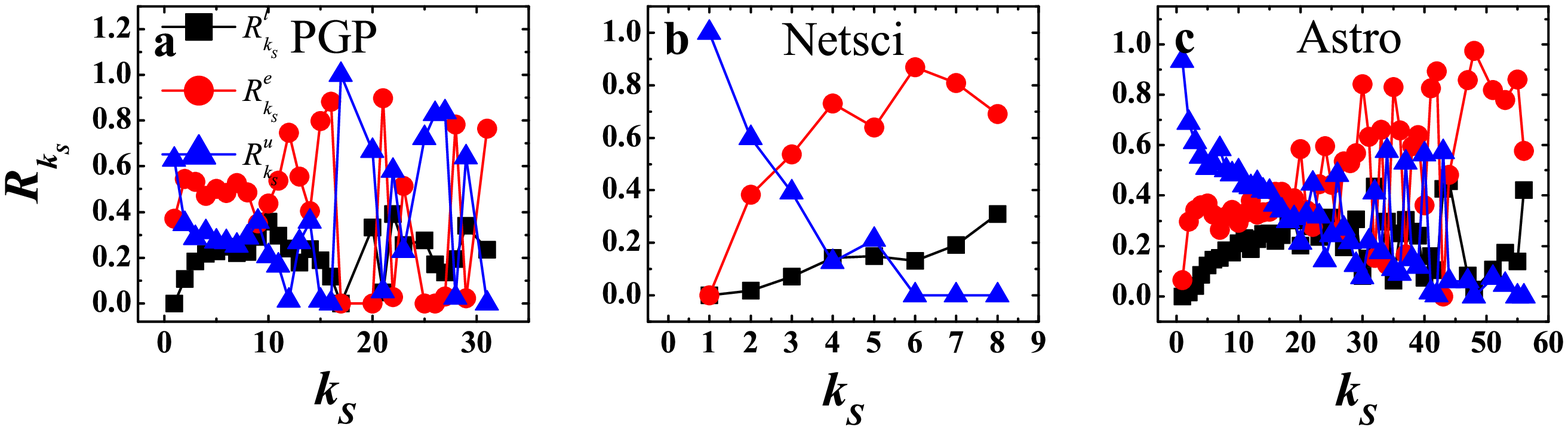,width=1\linewidth}
\renewcommand\thefigure{S\arabic{figure}}
\caption{\textbf{Link strength of shells for three real networks.} The link strength of each shell to its lower shells $R^{l}_{k_S}$ (black squares), equal shell $R^{e}_{k_S}$ (red circles) and upper shells $R^{u}_{k_S}$ (blue triangles) are represented. For those core-like groups, 21-shell, 16-shell and 15-shell in PGP, 7-shell and 6-shell in Netsci and 48-shell and 30-shell in Astro, $R^{e}_{k_S}$ is much higher than $R^{l}_{k_S}$.}
\label{figures8}
\end{center}
\end{figure}

\begin{figure}[!ht]
\begin{center}
\epsfig{file=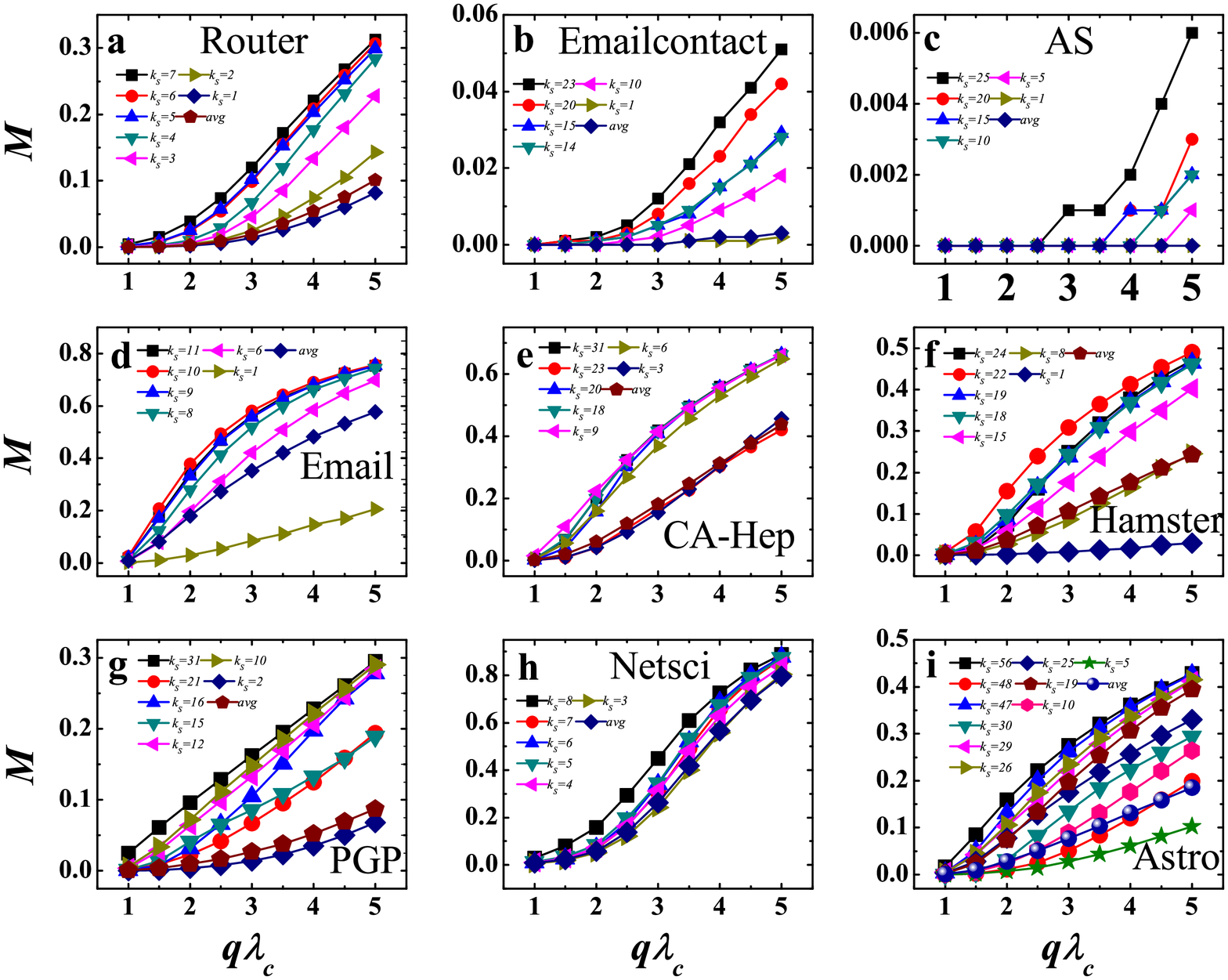,width=1\linewidth}
\renewcommand\thefigure{S\arabic{figure}}
\caption{\textbf{Infected population of $k_S$ shells as a function of infection probability, which is $q$ times of the epidemic threshold $\lambda_{c}$, $q$ ranges from 1 to 5.} (a), (b), (c) In the first group, high shells are consistently reaching a higher infection population than low shells. (d), (e), (f) In the second group, the innermost cores have a relatively low efficiency than low shells. (g), (h), (i) In the third group, the innermost cores have the highest spreading efficiency. But there exist some shells of high $k_S$ index that have a lower efficiency than adjacent shells. They are 21-shell, 16-shell and 15-shell in PGP, 7-shell and 6-shell in Netsci and 48-shell and 30-shell in Astro. }
\label{figures9}
\end{center}
\end{figure}

\begin{figure}[!ht]
\begin{center}
\epsfig{file=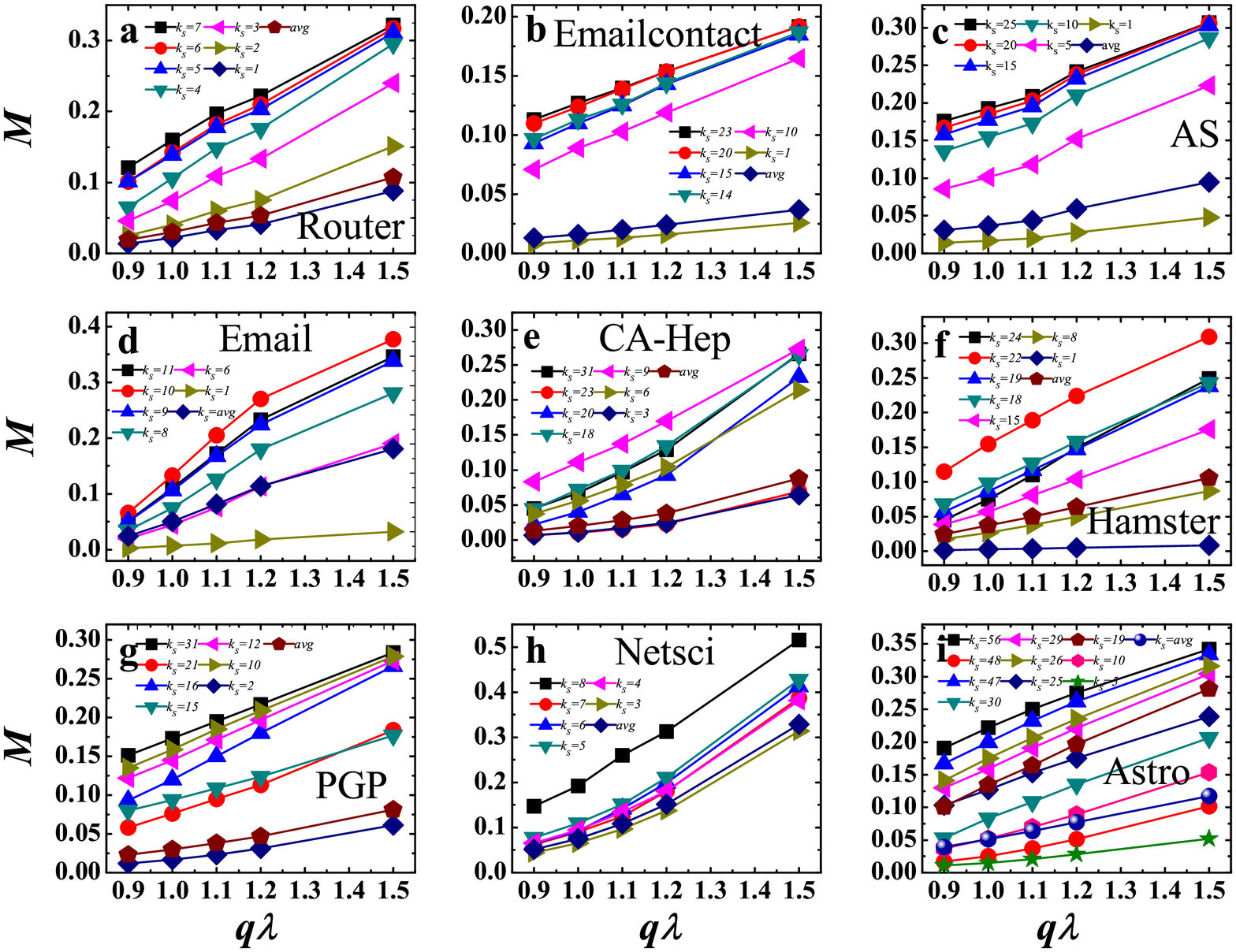,width=1\linewidth}
\renewcommand\thefigure{S\arabic{figure}}
\caption{\textbf{Infected population of $k_S$ shells as a function of infection probability, which is $q$ times of the infected probability $\lambda$, $q$ ranges from 0.9 to 1.5.} The relative spreading efficiency of shells is the same as the spreading when the infection probability is around $\lambda_{c}$ as shown in Figure S9.}
\label{figures10}
\end{center}
\end{figure}

\end{document}